\documentclass[preprint]{aastex}
\usepackage{color}
\usepackage[breaklinks, hyperindex]{}

\def\be{\begin{equation}}
\def\ee{\end{equation}}
\def \cage {\tau_{c} } 		
\def \mcage {\widetilde{\tau}'}	
\def \mage {\widetilde{\tau}}		

\newcommand{\dotline}{\protect\rotatebox[origin=lt]{160}{$...$} }
\newcommand{\dashbc}{\protect\rotatebox[origin=c]{160}{\small{${\bf- - }$}} }
\newcommand{\reddashline}{\protect\rotatebox[origin=c]{15}{\small{\textcolor{red}{- - -}}} }
\newcommand{\bluedashdot}{\protect\rotatebox[origin=c]{15}{\small{\textcolor{blue}{- $\cdot$ -}}} }
\newcommand{\ppd}{$P-\dot{P}$ } 
 
\newcommand{\ppdp}{($P-\dot{P}$) } 
\newcommand{\pdp}{($\dot{P}$) } 

\begin{document}

\shorttitle{Millisecond Pulsar Ages}
\shortauthors{K{\i}z{\i}ltan \& Thorsett}
\title{ {\sc Millisecond Pulsar Ages
\linebreak Implications of Binary Evolution and a Maximum Spin Limit
}}

\author{B\"ulent K{\i}z{\i}ltan \& Stephen E. Thorsett}
\affil{Department of Astronomy \& Astrophysics, University of California \& \\ UCO/Lick Observatory, Santa Cruz, CA 95064}
\email{bulent@astro.ucsc.edu}

\begin{abstract}

In the absence of constraints from the binary companion or supernova remnant, the standard method for estimating pulsar ages is to infer an age from the rate of spin-down. While the generic spin-down age may give realistic estimates for normal pulsars, it can fail for pulsars with very short periods. Details of the spin-up process during the low mass X-ray binary (LMXB) phase pose additional constraints on the period ($P$) and spin-down  rates \pdp that may consequently affect the age estimate. Here, we propose a new recipe to estimate millisecond pulsar (MSP) ages that parametrically incorporates constraints arising from binary evolution and limiting physics. We show that the standard method can be improved by this approach to achieve age estimates closer to the true age while the standard spin-down age may overestimate or underestimate the age of the pulsar by more than a factor of $\sim$10 in the millisecond regime.

We use this approach to analyze the population on a broader scale. For instance, in order to understand the dominant energy loss mechanism after the onset of radio emission, we test for a range of plausible braking indices. We find that a braking index of n=3 is consistent with the observed MSP population. We demonstrate the existence and quantify the potential contributions of two main sources of age corruption: the previously known ``age bias'' due to secular acceleration and ``age contamination'' driven by sub-Eddington progenitor accretion rates.  We explicitly show that descendants of LMXBs that have accreted at very low rates ($\dot{m}\ll \dot{M}_{Edd}$) will exhibit ages that appear older than the age of the Galaxy. We further elaborate on this technique, the implications and potential solutions it offers regarding MSP evolution, the underlying age distribution and the post-accretion energy loss mechanism. \footnote{Full resolution color figures and movies available at URL: http://www.kiziltan.org/research/MSP/ages.html}

\end{abstract}


	\section{\label{sec:intro}Introduction}

An accurate determination of pulsar ages plays a critical role in our understanding of advanced stages of stellar evolution, supernova explosions and remnants, white dwarf (WD) atmospheres and cooling models, binary evolution, planet formation around compact objects, and pulsar evolution in general. 

Typically, pulsar ages are estimated by calculating the amount of energy lost during their spin-down. Consequently, the spin-down age of a pulsar can be formulated as:
\begin{equation} 
\label{eq:age}
	\tau\; = \;  \frac{P}{(n-1)\;\dot{P}}\left[1-\left(\frac{P_0}{P}\right)^{n-1}\right] \\
\end{equation}
where the period ($P$) and the spin-down rate \pdp are the two main observables acquired by pulsar timing measurements. In the standard approach, the unknown initial spin period ($P_{0}$) of the pulsar is assumed to be much smaller ($P_{0}\ll P$) than the observed period. The dominant energy loss mechanism is analytically captured by the braking index, which is $n=3$ for pure dipole radiation and is implicitly adapted by the characteristic age $\tau_{c}$. The age of a pulsar can then be conveniently approximated by its characteristic age, where:
\begin{equation} 
\label{eq:cage}
	 \tau \longrightarrow \tau_{c}\equiv \frac{P}{2\;\dot{P}} \;\;{\rm for}\;\; P_{0}\ll P,\; n=3.
\end{equation}

While this approach may give reliable estimates for some normal pulsars (e.g. Crab Pulsar-PSR B0531+21: $\tau_{c}\sim 1240$~yr whereas the age from the supernova remnant (SNR) $\tau_{SNR}\sim 955$~yr; \citealp{WM77}), it should be kept in mind that characteristic ages for some other pulsars will suffer, dramatically in some cases (e.g. PSR~J0205+6449: $\tau_{c}\sim 5370$~yr, $\tau_{SNR}\sim 820$~yr; \citealp{MSS02}), because of the assumptions that render the standard approach less accurate, especially in the millisecond regime. Therefore, it is useful to develop a more comprehensive and detailed framework by which we can quantitatively understand MSP spin evolution.

In our previous work, we have set up a base by which we demonstrated that the proper inclusion of evolutionary constraints alone gives a deeper insight into the subsequent spin evolution \citep{KT09}.

In this paper, we propose a recipe to estimate pulsar ages that parametrically incorporates additional evolutionary and physical constraints. We show that the combined effect of a possible spin-up process during the LMXB phase and a maximum spin period due to the limiting centrifugal forces imparts meaningful constraints on the joint period ($P$) and spin-down \pdp values that MSPs can attain, which ultimately are used to estimate their ages. We detail the contribution this new approach offers to our understanding of MSP evolution and elaborate on the ramifications on several related problems such as the dominant energy loss mechanism and braking indices of millisecond radio pulsars, WD atmospheres and cooling models, the underlying age distribution, the enigma of MSPs that appear older than the galaxy they reside in, and the sources of MSP age corruption.

	\section{\label{sec:spin} The Spin Up Process}
	
	The period and spin-down \ppdp relation of a particular MSP at the epoch when it turns on its radio emission can be scaled as
\be
\dot{P} \propto \dot{m} P_{0}^{4/3}\,, 
\label{eq:spin}
\ee	
 as the neutron star can ultimately be spun-up no further than the spin period delineated by the Keplerian velocity at the Alfv\'en radius \citep{GL92}. We will use the least conservative upper bound of this scaling factor as an upper limit for the region where MSPs may be re-born. A single birth line instead would imply that these MSPs have accreted at near-Eddington rates ($\dot{m}\simeq \dot{M}_{Edd}$), which appears unreasonable, at least for a significant fraction of the observed MSP population as indicated by the paucity of sources near the spin-up line (see $\S$\ref{sec:cont} for discussion). 

We also now know that it is more than likely that the observed majority of MSPs do not initially re-appear in the vicinity of the spin-up line \citep{KT09}. The region where MSPs are re-born as radio sources on the \ppd diagram is strongly correlated with the dominant accretion rate ($\dot{m}$) experienced during the last phases of LMXB evolution, which is poorly constrained. For the scope of this paper, using the spin-up line as a marginal upper boundary for the region where MSPs are re-born as radio sources (i.e. where the spin-down trajectories on the \ppd plane start) will adequately account for possible nonlinear offsets due to other inherent uncertainties and assumptions made in Equation (~\ref{eq:spin}) regarding the accretion geometry (streams of hot plasma flowing onto the neutron stars' polar caps instead of uniform spherical or wind accretion) or the opacity of the accreted material (sensitive to whether the companion has an H or He rich envelope attached to a CO or ONeMg core, see $\S$:~\ref{sec:wd}).

On the other hand, the cumulative uncertainty of a presumed spin-up line that would offset or tilt a particular re-birth line cannot be arbitrarily big \citep{ACW99, FKR02}. Also, within the context of the standard recycling scenario \citep[see for review]{ACR82, RS82, BH91}, the ``spin-up line'' is merely an upper boundary below which MSPs are expected to be born rather than the line of culmination. To this day, there are no recycled pulsars observed above the spin-up line, except few in globular clusters (GCs) which have very uncertain evolutionary histories. Therefore, we will limit ourselves to Galactic MSPs whose spin-down history, orbital dynamics and Galactic kinematics remain unperturbed by gravitational encounters.
	
	\section{\label{sec:age} Millisecond Pulsar Ages}
		
The standard approach to estimate pulsar ages in the absence of additional constraints from either a possible association to an SNR or a stellar companion, has been to use the characteristic age as a proxy to the true age. The main goal of our work is to better understand the non-trivial relationship between the physically important true age and the observationally accessible characteristic age. We will refer to the time that has passed since the cessation of accretion as the ``(true) age: $\tau_{t}$'' of a recycled pulsar.

		\subsection{\label{sec:altmeth}Alternative Methods to Estimate Pulsar Ages}
		
			\subsubsection{\label{sec:kin}Ages From Kinematics and Supernova Remnants}
For some young pulsars that have reliable proper motion and distance measurements, a kinematic age estimate can be made by tracing the pulsar's trajectory in the galactic gravitational potential. However, without a firmly established birth site, kinematic ages are at best an indirect means to constrain pulsar ages. 

For MSPs, that have no associated SNRs and have ages that are much longer than the orbital timescales in the Galactic potential, all kinematic age information has been lost.

			\subsubsection{\label{sec:wd}White Dwarf Cooling Ages}
			
After the discovery of optical emission from pulsar companions \citep{K86}, WDs were soon realized as an alternative means to estimate the age of an MSP. Once active accretion has ceased in LMXBs with a recycled pulsar primary, the WD begins to subsequently radiate its internal heat after it burns off the remaining envelope. So, the beginning of WD cooling also marks the epoch when the spin-down starts for its companion. Therefore, in principle, the WD cooling ages are expected to be consistent with the ages of their MSP companions \citep{HP98.1, HP98.2}. 

The basics of WD cooling models are potentially accessible to theoretical understanding because of the simple thermal structure of the WD. The whole system is kept isothermal due to the efficient heat conduction of degenerate electrons. However, in practice, cooling ages remain difficult to estimate once realistic (and complex) effects of surface physics and stellar structure are included, leading to discrepancies and controversies in the interpretation of specific observations \citep{W92, SAM98, SEG00,  SDB00, ASB01.1, ASB01.2, KBJ05}. 

Possibly more promising than the original goal of using WD cooling ages to constrain the properties of MSPs, one might instead hope to use better constrained MSP ages ($\widetilde{\tau}$, see $\S$:~\ref{sec:rea}) to understand WD atmospheres and cooling models. 

\subsection{\label{sec:cha} Characteristic Ages: Idealized Pulsar Spin-down}

For millisecond radio pulsars, unbiased\footnote{For consistency, we designate unbiased values that are corrected for secular acceleration (i.e., Shklovskii effect) by adding `` $'$ '' to the parameter in lieu of referring them as ``intrinsic'' values. While the unbiased spin-down rates will represent the intrinsic values (i.e., $\dot{P}'=\dot{P}_{i}$), the unbiased characteristic age $\tau_{c}'=P/ 2\dot{P}'$ is neither the intrinsic nor the true age (i.e., $\tau_{c}'\ne \tau_{i}=\tau_{t}$, see \S~\ref{sec:cont} for discussion).
} 
characteristics ages ($\tau_{c}'$) become upper limits to the true ages ($\tau_{t}$) as the spin-down trajectories are truncated below the spin-up line. MSPs can be re-born on the spin-up line only if  they accrete at an Eddington rate ($\dot{m}=\dot{M}_{Edd}$) during recycling (see Equation (~\ref{eq:spin})). Because of likely sub-Eddington accretion rates experienced during the LMXB phase \citep{KT09}, the majority of the spin-down trajectories start well below the spin-up line. A considerable fraction of millisecond radio pulsars ($\sim$30\%) would even be expected to be born below the Hubble line (see $\S$:~\ref{sec:cont} for discussion). In this approximation, $\tau_{c}$ is derived by implicitly assuming pure dipole spin-down in the absence of other forms of additional torques and braking that might affect the apparent age. Other potential spin-down torques during the early stages after recycling such as gravitational or multipole radiation \citep{B98, K91} are also assumed to be absent when $\tau_{c}$ is inferred from the observed \ppd values. Possible non-monotonic field decay before magnetic stability sets in may also contribute to the corruption of characteristic ages. 

One bias for which we can properly correct is the effect of secular acceleration, i.e., Shklovskii effect \citep{S70}. The observed $\dot{P}$ values include an additional apparent spin-down factor introduced because of the increasing projected distance between the pulsar and the solar system barycenter. This leads to a quadratic centrifugal term \citep{CTK94},
\begin{eqnarray}
\label{eq:pc}
	\dot P_{s}= 1.08\times 10^{-18} \times \left( \frac{v_{t}} {100} \right)^{2} \times D_{kpc} ^{-1 }\times P
\end{eqnarray}
in
\begin{equation}
\label{eq:pdot}
\frac{\dot{P}}{P} \approx \frac{ \dot{P}' }{P} + \frac{v_{t}^2}{c D} \equiv \frac{\dot{P}' }{P} + \frac{\dot{P}_{s} }{P}
\end{equation}
where $\dot{P}$ and $\dot{P}'$ are the measured and unbiased spin-down rates for a pulsar at a distance $D$ (in units of kpc) with a transverse velocity $v_{t}$ (in units of 100 km s$^{-1}$).

Figure~\ref{fig:obs} shows the observed MSP population and the extent of the bias introduced by secular acceleration. MSPs that have a combination of relatively high transverse velocities and small distance measurements, have higher corresponding correction factors, e.g. for PSR J0034\(-\)0534 (D=0.98 kpc, $v_{t}\simeq$146.3 km s$^{-1}$) and PSR B1257+12 (D=0.77 kpc, $v_{t}\simeq$350.6 km s$^{-1}$) the corresponding correction factors are $\dot{P}/\dot{P}'\sim$9.3 and 16.9 respectively.

		\subsection{\label{sec:rea} A Realistic Age for Millisecond Pulsars ($\widetilde{\tau}$)}

One of the reasons why characteristic age estimates become less reliable for MSPs, even after the observed spin-down rates are unbiased for secular acceleration, is due to the assumption that the birth periods are much smaller than their currently observed spin periods (P$_{0}\ll$P), which fails for a considerable fraction of the population  (see $\S$~\ref{sec:cont}). \footnote{See a time-lapse movie for the true age evolution of millisecond pulsars at URL: http://www.kiziltan.org/research/MSP/ages.html}

In fact, the predicted MSP spin-down age is proportional to the integrated (and normalized) spin-down path from P$_{0}$ to P. The difference in the integrated  trajectories will depend on the initial spin period P$_{0}$, which remains an unknown parameter in most cases. We know that the spin-down trajectory in the case when recycling and a constraining maximum spin limit are taken into account is more compressed than what the standard approach predicts. Therefore, with tighter upper limits on the true age, $\widetilde{\tau}$ will give an age estimate that is closer to $\tau_{t}$.

Given enough time, MSPs may reach a limiting equilibrium phase where they begin to lose angular momentum gained from accretion by shedding mass. A maximum spin limit beyond which a neutron star cannot be spun-up due to this continuous loss of excess angular momentum will truncate the spin-down trajectories vertically. Several authors \citep{HZ89, FI92, CST94} have constructed equilibrium sequences for neutron stars where a range of mass shedding periods are calculated for different equations of states including the effects of rapid rotation and large deviations from spherical symmetry. While the theoretical best-fit values range between P$_{sh}=$1.28-1.32~ms for neutron stars with realistic configurations, for some extreme cases they find that a limiting period of P$_{sh}\gtrsim$0.85 ms may be plausible. \cite{CMM03} find evidence for a statistically significant upper limit at P$_{sh}\simeq$1.32~ms. We therefore use a putative value of P$_{sh}\sim$1~ms and include it parametrically to perform calculations.

The critical magnetic field B$_{c}$ will be the locus of points extending from the intersection of the diagonal spin-up line and the vertical mass shedding limit. MSPs with magnetic fields below B$_{c}$ can only be spun-up to the limiting mass shedding period. 
The critical magnetic field is B$_{c}$= (3.36$^{+0.7}_{-0.9}$)$\times10^{8}$ G for P=1 ms where the spin-up line is prescribed as $\dot{P}=\alpha P^{4/3}$ for $\alpha=(1.1\pm0.5)\times10^{-15}${\bf s}$^{-4/3}$ \citep{ACW99}.

We can formulate an age estimate that implements the constraints arising from recycling and mass shedding:
\begin{eqnarray}
	\widetilde{\tau}\,(B > B_{c})&=&\frac{P}{(n-1)\;\dot{P}}\left[1-\left(\frac{\widetilde{\alpha}\, 
		\dot{P}^{3/7}}{P^{4/7}}\right)^{n-1}\right]\label{eq:mage1}			\\
	\widetilde{\tau}\, (B < B_{c})&=&\frac{P}{(n-1)\;\dot{P}}\left[1-\left(\frac{P_{sh}}{P}\right)^{n-1}\right]\label{eq:mage2}
\end{eqnarray}
where P$_{sh}$ is the mass shedding limit. We parametrically adopt the re-normalized coefficient $\widetilde{\alpha}=2.6^{+0.7}_{-0.4} \times10^{6}{\bf s^{4/7}}$ and use it as a fiducial value. Although $\widetilde{\alpha}$ inherently has numerous sources of uncertainty, the corresponding minimum post-accretion period that defines the spin-up line depends on the total accreted mass, but is insensitive to other parameters \citep[see][]{PK94}. \cite{ACW99} find an upper limit of $\alpha=1.6\times10^{-15}${\bf s}$^{-4/3}$ to be empirically reasonable which we adopt to marginalize $\widetilde{\alpha}$.
	
At low magnetic fields (B$\lesssim$ B$_{c}$), the Alfv\'en radii ($r_{A}\propto B_{s}^{4/7}$) shrink essentially down to the surface of the neutron star and angular momentum is transferred more efficiently and stably on longer timescales.	Therefore the spin-down trajectories for neutron stars with lower magnetic fields are more likely to start closer to the vertical re-birth line ($\sim$P$_{sh}$). 

The level of truncation of the spin-down trajectories can be more dramatic for neutron stars with very low magnetic fields and short periods (P$\lesssim3$ ms). These MSPs with B$\lesssim10^{8}$G are expected then to be born with periods very close to their observed ones (P$\simeq$P$_{0}$) and therefore are more likely to be much younger than they appear.

\begin{figure} [t!]
	\includegraphics[width=4.9in,angle=90.]{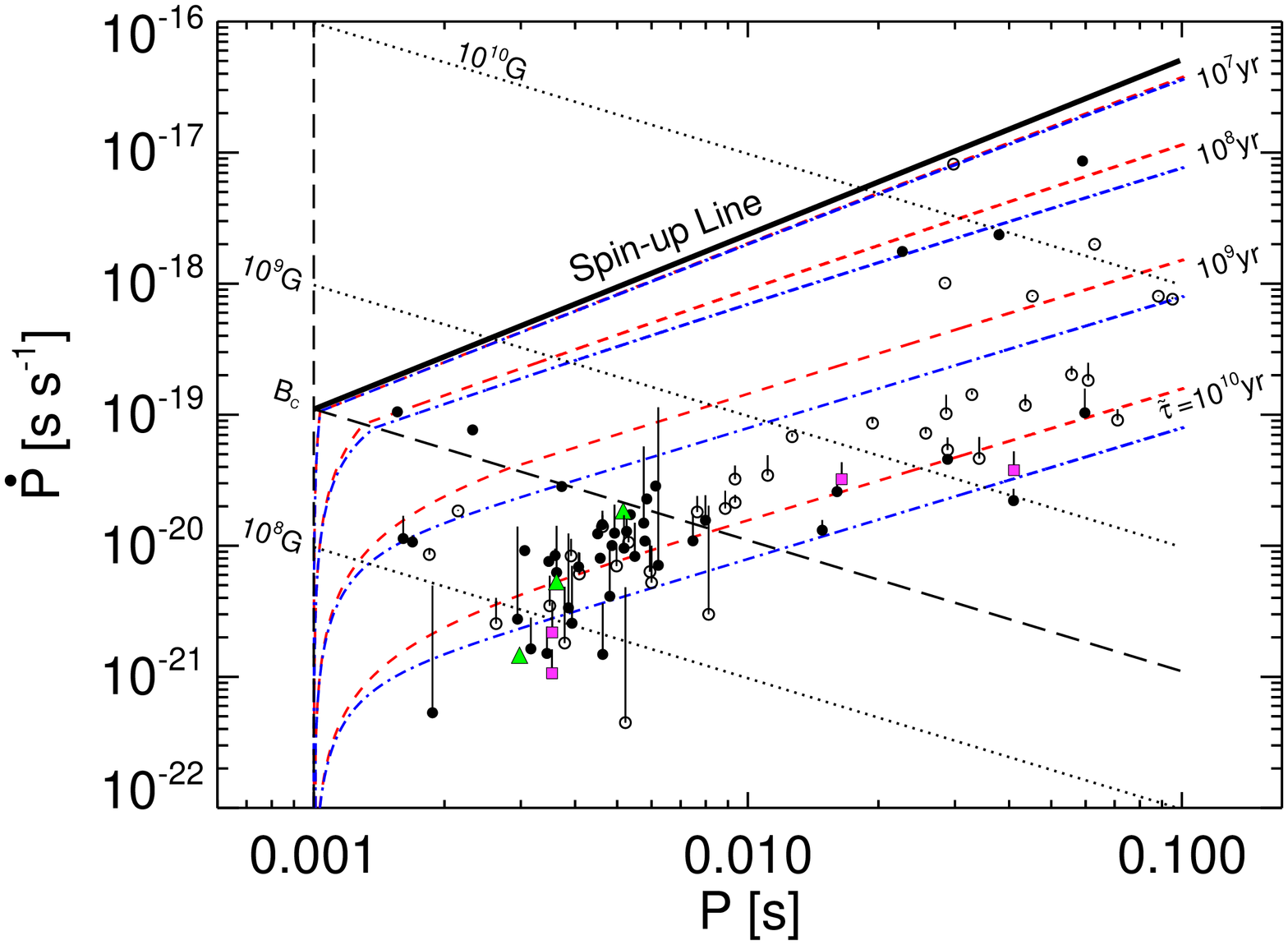}
	\figcaption{Observed distribution of Galactic MSPs: black dots ($\bullet$) indicate unbiased spin-down rates ($\dot{P}'$) for MSPs that have proper motion measurements. Vertical solid lines are the corresponding Shklovskii correction. MSPs with no proper motion measurements ($\circ$) are corrected for an assumed $v_{t}=100\,$km s$^{-1}$ in order to see the potential level of bias. The magenta squares are MSPs (\textcolor{magenta}{$\blacksquare$}: PSRs J1022+1001, J1216--6410, J1829+2456 and J1933--6211) for which the spin-down rates are corrected only for an assumed $v_{t} = 50\,$km s$^{-1}$ while these MSPs will otherwise appear to be spinning-up, i.e. $\dot{P}'_{v_{100}} < 0$. The green triangles are MSPs (\textcolor{green}{$\blacktriangle$}: PSRs J1024--0719, J1801--1417 and J2229+2643) with proper motion measurements for which $\dot{P}'< 0$ and therefore are left uncorrected. The red and blue lines are the MSP age ($\mage$) lines for braking indices n=3 (\reddashline) and 5 (\bluedashdot). The diagonal solid line is the spin-up line $\dot{P} \sim \dot{m} P_{0}^{4/3}$ for $\dot{m}=\dot{M}_{Edd}$. The diagonal dotted lines (\dotline) are the inferred dipole field strengths. $B_{c}$ (\dashbc) is the critical magnetic field below which MSPs can be spun up to the mass shedding limit $P_{sh}\simeq$1 ms.	\label{fig:obs}    }	
	\end{figure}

In Figure~\ref{fig:obs} the sets of blue and red lines are MSP age ($\mage$) lines for braking indices n=3 (red: dash) and 5 (blue: dash-dot). It is noteworthy to point out that for MSPs whose angular momentum and energy loss is dominated by multipole or quenched by gravitational wave radiation, the scaling factor in Equation (~\ref{eq:spin}) can be different. The blue line in Figure~\ref{fig:obs} demonstrates the potential level of additional bias for the case in which gravitational wave radiation is the dominant mechanism for energy loss as opposed to braking due to pure dipole radiation. MSPs that lose energy via more efficient processes will traverse the spin-down path much faster, and consequently mimic older ages. As a result, the contribution of more efficient processes will exacerbate the age overestimate further.

\begin{figure*}[t]
	\includegraphics[width=4.9in,angle=90]{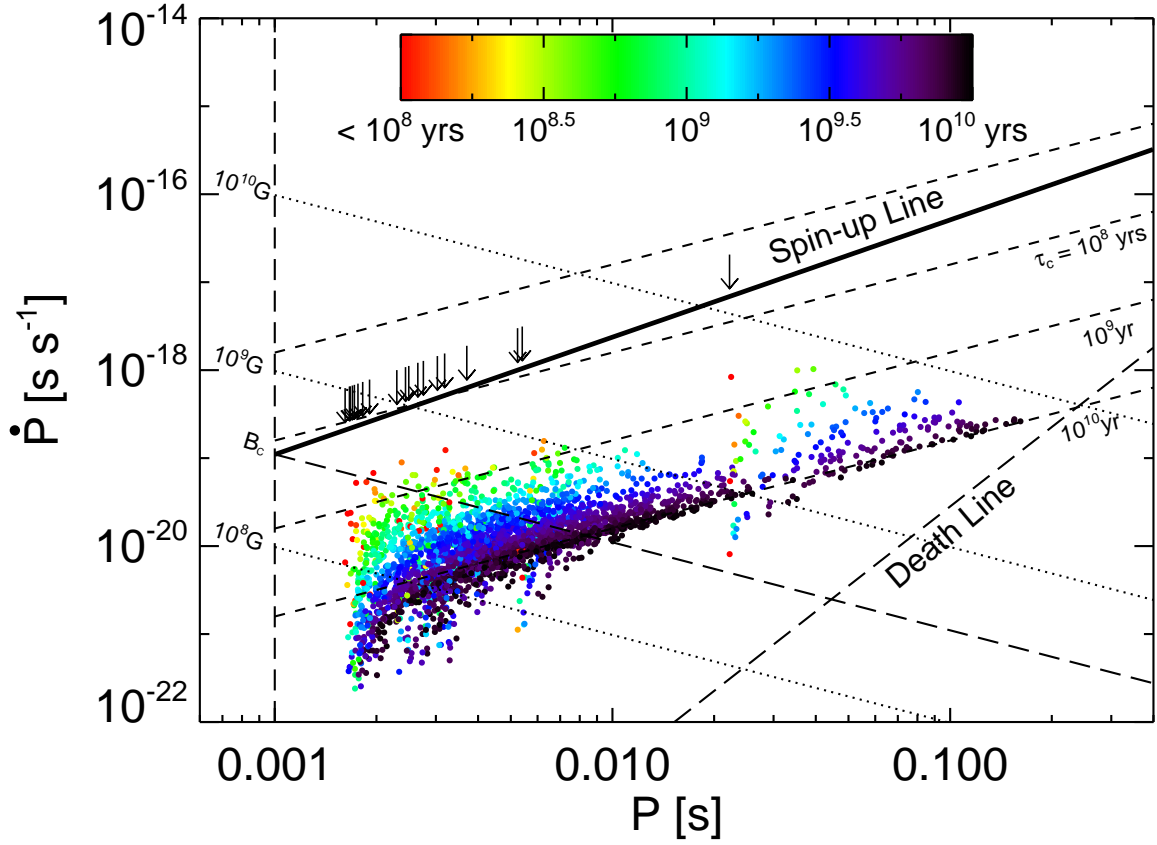}
	\figcaption{Expected true age distribution of the underlying MSP population:  color represents the true age ($\tau_{t}$). Downward arrows ($\downarrow$) are the neutron star spin frequencies measured in LMXBs which are used as progenitor seeds to reconstruct the synthetic population.
\label{fig:sim}   
   }	
	\end{figure*}

Figure~\ref{fig:sim} shows the true age trend for MSPs. The synthetic population is produced with the method described in \cite{KT09}, where the MSP evolution is parameterized by the {\tt evolution functional} $\mathcal{E}(D, \dot{M}, R)$, in which $D$ is the initial period (P$_{0}$) distribution of the progenitor population, $\dot{M}$ is the distribution of predominant accretion rates experienced during the latest phases before the onset of radio emission, and $R$ is the galactic birth rate. The parameter space ($D, \dot{M}, R$) is sampled by producing purely random synthetic populations. Then, only the input parameters that produce synthetic samples consistent with the observed MSP population  are used for the construction of the underlying \ppd demographics. For the filtration process, we use a {\tt {\em relaxed} multi-dimensional Kolmogorov-Smirnov (K-S) test} as the consistency criteria for the multi-layered Monte-Carlo integration scheme \citep[and references therein]{FF87}.

We choose sub-samples that are uniformly selected from the underlying MSP population to produce Figures~\ref{fig:sim} and~\ref{fig:multi}. While selection biases play a role in which MSPs are preferentially observed, the reflected age trend will remain unaffected. Therefore, the age trend of the underlying population both in Figures~\ref{fig:sim} and~\ref{fig:multi} is expected to be a realistic reflection of the true age.
	\section{\label{sec:dis}Discussion}
	
		\subsection{\label{sec:cont}Sources of Age Corruption: Bias and Contamination}

Older MSPs tend to appear younger if the spin-down rates ($\dot{P}$) are not properly corrected for the contribution due to secular acceleration. This causes an upward ``age bias'' on the \ppd plane. The Shklovskii effect is more pronounced for MSPs that are closer to the solar system barycenter and have higher transverse velocities.

We can unbias our measurements and correct for this corruption by calculating the centrifugal term (Equations (\ref{eq:pc}) and (\ref{eq:pdot})) once both the distance and transverse velocity terms are accurately known. The proper motion measurements for MSPs with poorly constrained distances pose a stiffer challenge than previously predicted \citep{DTB09}. To avoid underestimating the potential age bias, we introduced conservative random deviations to the observed distance and transverse velocity measurements by up to $\pm 0.4D$ and $\pm 0.05 v_{t}$ \citep{BBG02, BFG03, DTB09} to produce the sample population in Figure~\ref{fig:multi}. 

Figure~\ref{fig:multi}(a) reflects the potential age bias we estimate for the underlying population. In some cases, relatively old MSPs may even appear above the spin-up line if the observed spin-down rates are not properly unbiased. The age bias caused by the Shklovskii effect will tend to push older MSPs upward by amounts proportional to their correction term ($\dot{P}/\dot{P}'$) and hence make them appear younger.

Throughout this work, we have excluded MSPs in GCs due to their complicated spin evolution. 
The cluster's compact gravitational well, and to a lesser extent the larger cross section for gravitational interaction in GCs relative to the Galactic disk, effectively perturb the spin evolution of these MSPs by injecting a cumulative corruption to their spin-down rates. Conversely, the spin-down trajectories of MSPs in the Galaxy will include less corrupted information about the initial formation. Even though the level of age bias for Galactic MSPs appears not as dramatic as the ones that are gravitationally perturbed in GCs, it still mixes the apparent age distribution quite efficiently (Figure \ref{fig:multi}(a)). 

Unbiasing will push old MSPs that appear younger to their corresponding age ($\mage$) lines. This correction process will recover the final positions of the true spin-down trajectories but will effectually exacerbate the downward age contamination which was artificially diluted by secular acceleration. The main source of the downward age contamination though will be the sub-Eddington progenitor accretion rates experienced during the LMXB phase. 

These MSPs are expected to be born with smaller spin-down rates and consequently traverse shorter trajectories. We predict that $\sim$30\% of MSP ages overestimate the true age by more than a factor of 2. Therefore, we argue that MSPs, which were presumed as the oldest sub-population, have a flatter true age distribution than previously thought.

While we can reverse the upward ``age bias'', the downward ``age contamination'' on the other hand, reflects upon the intrinsic spin-down rates and is real  (Figures~\ref{fig:multi}(a) and (b)). Unless we have a unique leverage to accurately determine the period at birth (P$_{0}$) along with the respective progenitor accretion rate, the inferred MSP ages will remain to be strict overestimates (Figure~\ref{fig:multi}(b)).

\begin{figure*}
		\includegraphics[width=4.9in,angle=90]{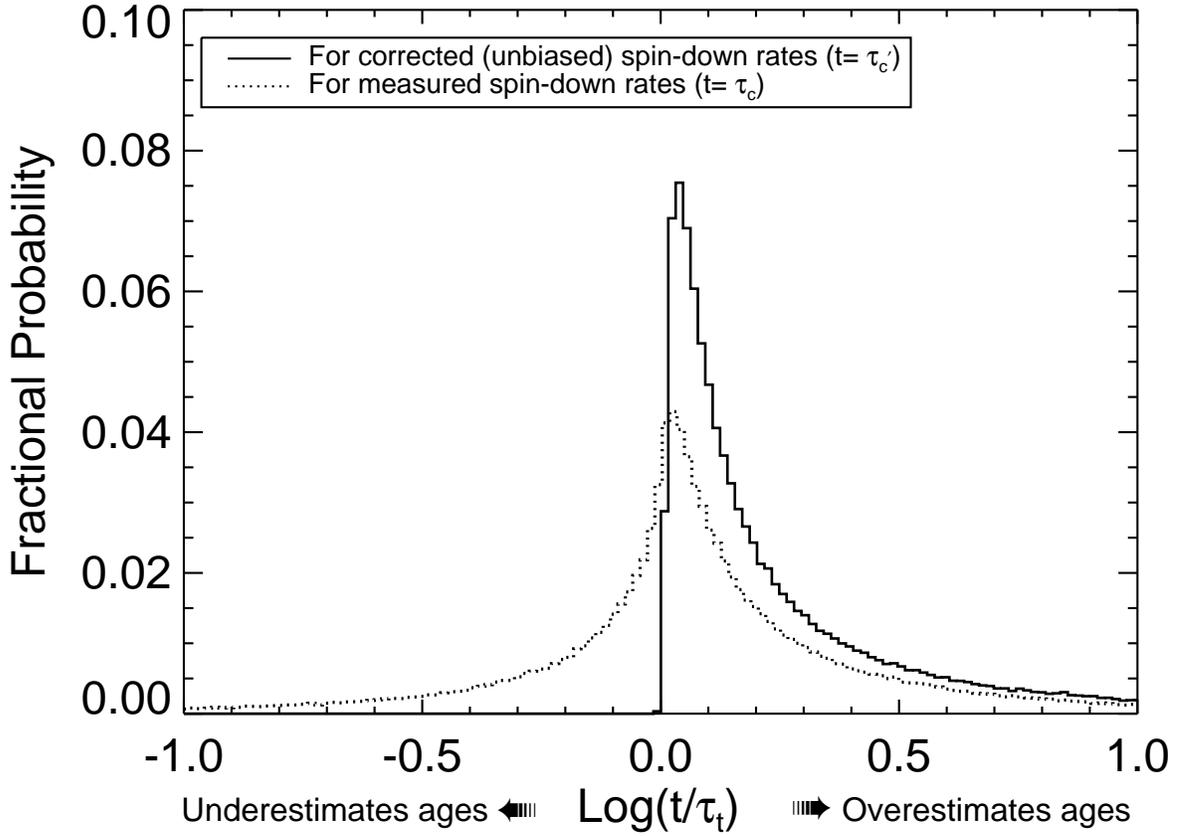} 	
	  	 \caption{Effects of secular acceleration on the MSP age distribution (see also  Figures~\ref{fig:multi}): ages inferred from measured $P$ and $\dot{P}$ values will overestimate or underestimate the true age (dashed line). Unbiased characteristic ages ($\tau_{c}'$) reflect mere overestimates (solid line) where $\sim$30\% of MSPs will have a $\tau_{c}'$ overestimating the true age by more than a factor of 2.}
		 	\label{fig:dist}   
\end{figure*}

Figure \ref{fig:dist} quantifies the level of bias and contamination of MSP ages before and after correcting for the contribution due to secular acceleration. Both the upward bias and downward contamination will simultaneously remain before the spin-down rates are properly corrected for the Shklovskii effect. We predict that 20\% of the measured \ppd values for MSPs (dotted line in Figure~\ref{fig:dist}) will overestimate the age by more than a factor of two, whereas about 10\% will underestimate the age at the same level. Old and young MSPs are practically indistinguishable before the spin-down rates are properly unbiased. Some very young MSPs may appear even below the Hubble line whereas much older MSPs may be observed above the spin-up line (Figure~\ref{fig:multi}(a)). The asymmetric wings of the dotted line that extend in both directions in Figure~\ref{fig:dist} quantify this bidirectional corruption.

The ages obtained from intrinsic \ppd values will represent strict upper limits to the true age  ($\tau_{i}=\tau_{t} \le \widetilde{\tau}'\le \tau_{c}'$). The truncated solid line in Figure \ref{fig:dist} shows at what level we overestimate the ages for the whole MSP population. After properly unbiasing the observed spin-down rates, we expect to overestimate the age of 30\% of MSPs by more than a factor of two. Table 1 shows $\cage$ and $\mage$ for observed millisecond radio pulsars before and after unbiasing. Table 2 marginalizes the potential bias for an assumed $v_{t}\sim$100 km s$^{-1}$ for millisecond radio pulsars that have no proper motion measurements.

The intrinsic \ppd values of the underlying MSP population suggest that $\sim$30\% of MSPs will be born with apparent ages older than the age of the Galaxy. The true age distribution of MSPs with $\tau_{c}\ge 10^{10}$ yr is relatively well mixed as opposed to MSPs that lie on or just above the Hubble line. Hence, the sources that appear below the $10^{10}$yr line might not be among the oldest within the  MSP sub-population.


We predict younger ages for pulsars with $\tau_{c}/\mage'>$1 (Table 1). The majority of MSPs for which we predict younger ages, are the ones that are closer to the spin-up line. These MSPs have a significant fraction of their spin-down trajectories truncated because the timescales pulsars spend close to the spin-up line is much shorter than MSPs with smaller spin-down rates. Some of the younger sources with $\tau_{t}\le\mage'<\cage$ are PSRs J0218+4232, J0737\(-\)3039A, J1023+0038, B1534+12, B1913+16,  and B1937+21.

For MSPs with no proper motion measurements, Table 2 shows the potential biases for an assumed $v_{t}$=100 km s$^{-1}$. In this category, PSR J1841+0130 may have the strongest age corruption with $\cage/\mage'\sim$5.85.

There are two sources in particular that have been of considerable interest:

{\em PSR J1012+5307}: The ranges of cooling ages for possibly the best studies example is PSR J1012+5307 with $\tau_{wd}\sim$0.3-7~Gyr \citep{LL95, ASH96, BKW96}. We derive an age of $\widetilde{\tau}'\sim$ 6.25~Gyr (Table 1) for PSR J1012+5307 which is consistent with $\tau_{c}'$. This implies that the true age has to be $\le$ 6.25 Gyr. One cannot exclude younger ages by either the Shklovskii corrected (unbiased) characteristic age ($\tau_{c}'$) or MSP age ($\widetilde{\tau}'$) approach as PSR~J1012+5307 may have been born with a P$_{0}$ very close to its currently observed period (P$_{0}\simeq$P) due to low accretion rates experienced during the LMXB phase (see $\S$\ref{sec:spin}). 
		
		
{\em PSR B1257+12}: The cumulative correction to the age of PSR B1257+12 is among the most significant of MSPs that have proper motion measurements ($\dot{P}/\dot{P}'\sim$16.6). We infer an upper limit of $\tau_{t}\le\mcage=1.42\times10^{10}$~yr for the age. This implies that the age of PSR B1257+12 cannot be constrained solely from its spin-down history. 

In general, when interpreting ages inferred from spin-down histories for single MSPs, one has to consider the possibility that an evolutionary process which produces MSPs without a binary companion may affect the spin-down evolution. For instance, an encounter by which the companion is ejected will perturb the spin-down rate of the compact primary. In the exceptional case of PSR B1257+12, the process that led to the formation of the planets may have affected the spin-down evolution.

 
		\subsection{\label{sec:bra} Braking Index}

We tested whether alternative energy loss mechanisms other than pure dipole braking (n=3) are required to reconcile for the MSP age distribution. While more efficient processes (n$>$ 3) may also contribute to the downward age contamination, it would seem unnecessary to invoke higher braking indices to account for ages that appear older than the Galaxy. We do not rule out that gravitational wave radiation may expedite spin-down during the very early stages after re-birth before magnetic stability sets in \citep{L95, LM95, B98}. However, for MSPs following standard spin-up, the contribution of gravitational wave radiation to age contamination may not to exceed the offset between the blue-to-red $\mage$ age lines in Figure~\ref{fig:obs}. We explicitly show that lower preferred accretion rates during the active accretion phase can produce the paradoxically older appearing MSPs (Figures~\ref{fig:sim} and \ref{fig:multi}). 

Based on the \ppd characteristics of MSPs, we find no compelling evidence that energy loss has been  dominantly driven by multipole or gravitational wave radiation during a significant portion of the lifetime of these sources. 
 
 \section{\label{sec:con}Conclusions}
 
We have implemented constraints arising from the spin-up process and a limiting maximum spin limit into the standard method to obtain a more realistic age  ($\mage$) estimate for MSPs. There are a range of ramifications that follow:

{\em Age distribution}: The unbiased characteristic ages are only upper limits to the true age. The new age estimate gives tighter upper limits and hence is closer to the true age ($\tau_{t}\le\mcage\le\cage'$). This flattens and shifts the age distribution toward younger ages while the age corruption scrambles the positions on the \ppd plane quite efficiently. We predict that a significant fraction of MSPs are born with apparent older ages. 
The true age distribution of MSPs does not appear to peak at $\sim10^{10}$yr as sharply as expected for a sub-population recycled from a first generation of pulsars with features indicative of a population that is already old, at least in a dynamic sense \citep{HP97}. MSPs that appear older than the Galaxy can be reconciled with very low ($\dot{m}\ll\dot{M}_{Edd}$) progenitor accretion rates experienced during the latest phases of the LMXB evolution. We expect $\sim$30\% of the population to be born with $\tau_{c}'\ge 10^{10}$ yr.

{\em Age corruption}: There are two sources of age corruption: (1) the previously known ``age bias'', which appears to be more prominent than previously predicted and (2) ``age contamination'' which is driven by lower progenitor accretion rates. Age contamination, which effectively disguises young MSPs as old ones, is not correctable in the absence of additional constraints that may give us insight into the details of individual prior accretion histories. On the other hand, the correctable age bias, manifests itself as reverse contamination and will disguise old MSPs as younger sources. The downward contamination will remain as the main source of confusion with regards to MSP ages. We expect to overestimate the true age of MSPs by more than a factor of 2 for $\sim$30\% of the population. As a consequence, the birth and merger rates of NS--NS systems based on $\cage$ are most likely underestimates which therefore will have ramifications for potential LIGO sources.
 
{\em Braking indices}: The millisecond radio pulsar demographics is consistent with the canonical spin-down model (n=3). The challenge to disentangle possibly mixed sub-populations of MSPs that may have experienced dissimilar energy loss histories (n$\le$3 or n=5) is mainly due to the paucity of sources. Therefore, an early short phase when MSP energy loss is dominated by gravitational wave or multipole emission remains as a potentially contributing source of age contamination.  

\acknowledgements 
The research presented here has made extensive use of the 2009 August version of the ATNF Pulsar Catalogue \citep{MHT93}. BK thanks Athanasios Kottas for long discussions on Bayesian statistics which have seeded the idea and subsequently given birth to the proper parameterization of millisecond pulsar evolution. We thank the anonymous referee for useful comments. The authors acknowledge NASA and NSF grants AST-0506453.




\begin{figure}
		\includegraphics[width=6.4in,angle=0]{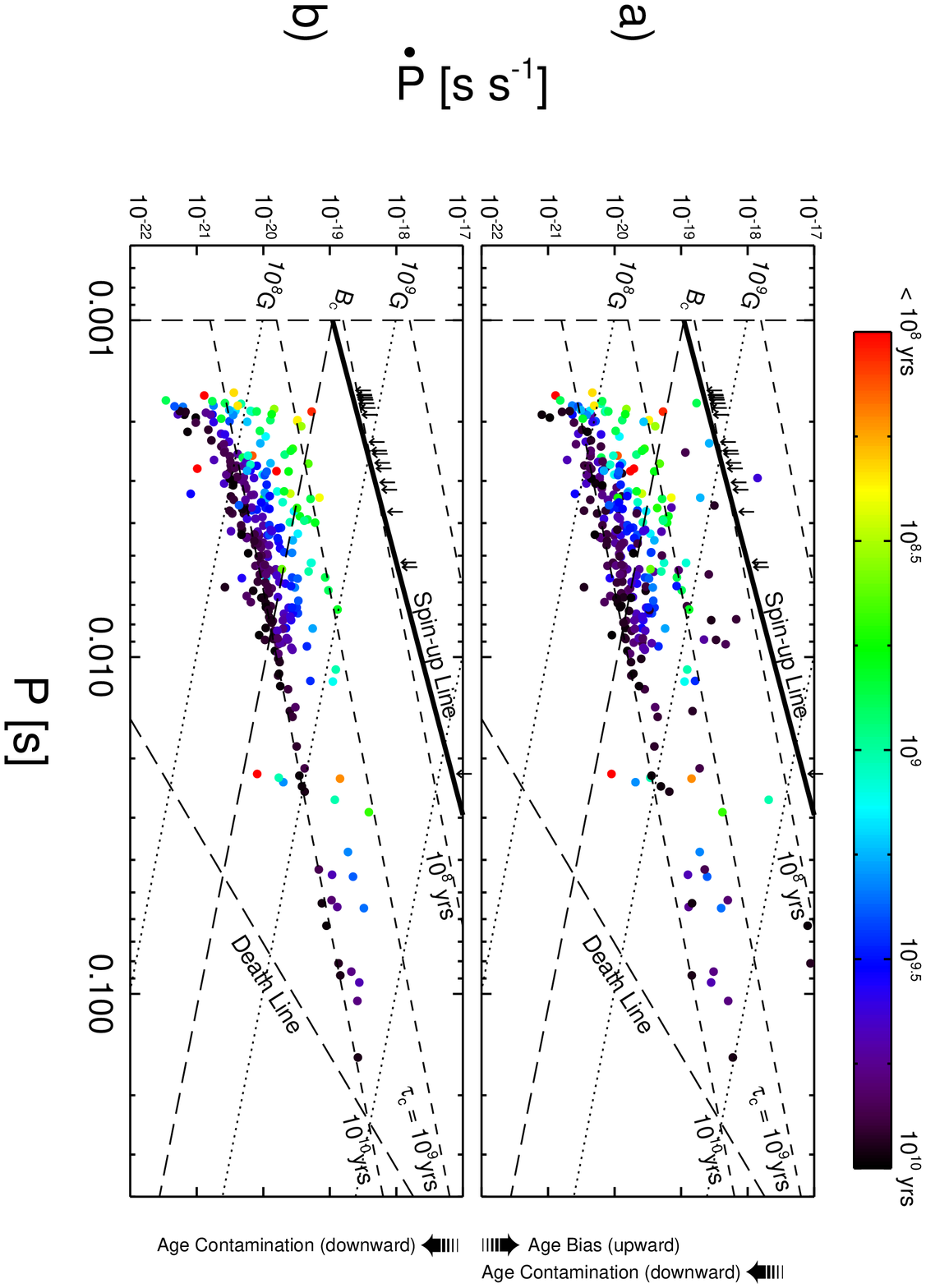} 	
		 	\label{fig:multi}   
\end{figure}

\begin{figure}
		  	 \caption{\ppd and underlying age distribution of MSPs (a) before and (b) after correcting for effects of secular acceleration: color represents the true age ($\tau_{t}$) and the diagonal dashed lines are the inferred characteristic age ($\tau_{c}$) lines. The progenitor seeds used to produce the synthetic population are randomly sampled from the period distribution of observed millisecond X-ray pulsars (downward arrows). {\bf (a)} The transverse velocities and distances are chosen to be consistent with the observed millisecond radio pulsar population in order to see the potential bias in the \ppd demographics. {\bf(b)} The \ppd distribution for the same sample MSP population after correcting for effects of secular acceleration. The process that disguises older MSPs as younger sources can be reversed by properly correcting (unbiasing) for the Shklovskii effect, whereas the population will still harbor young MSPs with apparent older ages. This (downward) age contamination is driven by the lower mass accretion rates experienced during the LMXB phase. We predict that a significant fraction of millisecond radio pulsars ($\sim$30\%) will be born with characteristic ages older than the age of the Galaxy.		 		 }
		 	\label{fig:multi}   
\end{figure}

\begin{deluxetable}{lccccc}
\label{tab:1}
\tablecolumns{4}
\tabletypesize{\scriptsize}
\singlespace
\tablewidth{0pt}
\tablecaption{Ages of millisecond pulsars with proper motion measurements.}
\tablehead{ 
\colhead{Pulsar} & 
\colhead{$\cage$ [Gyr]} &
\colhead{$\cage'$ [Gyr]} &
\colhead{ $\widetilde{\tau}$ [Gyr]} & 
\colhead{ $\widetilde{\tau}'$ [Gyr]} & 
\colhead{$\cage/ \widetilde{\tau}'$}
}
\startdata
    &       &         &    &     \\
 *  J0030+0451      &     7.56      &   7.65      &     7.24      &   7.32      &   1.03     \\
 *  J0034\(-\)0534      &     6.00      &  55.71      &     4.29      &  39.90      &   0.15     \\
 *  J0218+4232      &     0.48      &   0.48      &     0.34$_{-0.09}^{+0.04}$      &   0.35$_{-0.09}^{+0.04}$       &   1.37$_{-0.14}^{+0.48}$     \\
 *  J0437\(-\)4715      &     1.59      &   6.12      &     1.47      &   5.94      &   0.27     \\
 *  J0610\(-\)2100      &     4.93      &  17.92      &     4.60      &  16.72      &   0.30     \\
 *  J0613\(-\)0200      &     5.06      &   5.28      &     4.52      &   4.72      &   1.07     \\
 *  J0621+1002      &     9.67      &  10.01      &     9.56      &   9.91      &   0.98     \\
 *  J0711\(-\)6830      &     5.80      &  10.46      &     5.61      &  10.11      &   0.57     \\
 * J0737\(-\)3039A      &     0.20      &   0.20      &     0.14$_{-0.04}^{+0.02}$       &   0.14$_{-0.04}^{+0.02}$       &   1.43$_{-0.18}^{+0.57}$      \\
 *  J0751+1807      &     7.08      &   7.25      &     6.49      &   6.66      &   1.06     \\
 *  J1012+5307      &     4.87      &   6.48      &     4.69      &   6.25      &   0.78     \\
 *  J1023+0038      &     2.23      &   2.50      &     1.45      &   1.62      &   1.37     \\
 *  J1045\(-\)4509      &     6.77      &  10.91      &     6.63      &  10.71      &   0.63     \\
 *    B1257+12      &     0.86      &  14.58      &     0.75      &  14.21      &   0.06     \\
 *  J1453+1902      &     7.91      &   8.46      &     7.68      &   8.21      &   0.96     \\
 *  J1455\(-\)3330      &     5.21      &   8.07      &     5.07      &   7.93      &   0.66     \\
 *  J1518+4904      &    23.84      &  29.34      &    23.74      &  29.23      &   0.82     \\
 *    B1534+12      &     0.25      &   0.25      &     0.19$_{-0.04}^{+0.01}$      &   0.20$_{-0.04}^{+0.01}$      &   1.24$_{-0.05}^{+0.32}$     \\
 *  J1600\(-\)3053      &     6.00      &   6.76      &     5.54      &   6.24      &   0.96     \\
 *  J1603\(-\)7202      &    14.98      &  17.94      &    14.85      &  17.81      &   0.84     \\
 *  J1640+2224      &    17.71      &  30.59      &    15.94      &  27.53      &   0.64     \\
 *  J1643\(-\)1224      &     3.96      &   5.04      &     3.77      &   4.81      &   0.82     \\
 *  J1709+2313      &    20.21      &  49.45      &    19.27      &  47.15      &   0.43     \\
 *  J1713+0747      &     8.49      &   9.01      &     8.08      &   8.58      &   0.99     \\
 *  J1738+0333      &     3.85      &   4.07      &     3.71      &   3.93      &   0.98     \\
 *  J1744\(-\)1134      &     7.24      &   9.36      &     6.80      &   8.80      &   0.82     \\
 *    B1855+09      &     4.80      &   4.92      &     4.63      &   4.75      &   1.01     \\
 *  J1909\(-\)3744      &     3.34      &  17.08      &     2.95      &  15.12      &   0.22     \\
 *  J1911\(-\)1114      &     4.05      &   9.13      &     3.74      &   8.44      &   0.48     \\
 *    B1913+16      &     0.11      &   0.11      &     0.07$_{-0.03}^{+0.01}$      &   0.07$_{-0.03}^{+0.01}$      &   1.66$_{-0.29}^{+1.09}$      \\
 *    B1937+21      &     0.24      &   0.24      &     0.10$_{-0.09}^{+0.04}$      &   0.10$_{-0.09}^{+0.04}$       &   2.37$_{-0.66}^{+21.63}$     \\
 *  J1944+0907      &     4.80      &   8.59      &     4.63      &   8.27      &   0.58     \\
 *    B1953+29      &     3.27      &   3.41      &     3.14      &   3.27      &   1.00     \\
 *    B1957+20      &     1.51      &   2.23      &     0.92      &   1.37      &   1.10     \\
 *  J2019+2425      &     8.88      &  24.34      &     8.31      &  22.77      &   0.39     \\
 *  J2051\(-\)0827      &     5.63      &   5.81      &     5.35      &   5.52      &   1.02     \\
 *  J2124\(-\)3358      &     3.79      &   6.25      &     3.64      &   5.99      &   0.63     \\
 *  J2129\(-\)5721      &     1.98      &   2.09      &     1.84      &   1.94      &   1.02     \\
 *  J2145\(-\)0750      &     8.53      &   9.81      &     8.42      &   9.69      &   0.88     \\
 *  J2235+1506      &     5.99      &   9.13      &     5.92      &   9.04      &   0.66     \\
 *  J2317+1439      &    22.55      &  36.18      &    20.65      &  33.13      &   0.68     \\
 *  J2322+2057      &     7.85      &  18.49      &     7.51      &  17.69      &   0.44     \\
 \enddata
\tablecomments{PSRs~J1024\(-\)0719, J1801\(-\)1417 and J2229+2643 have $\dot{P}'< 0$ for $v_{t}$=146, 208 and 115\,km s$^{-1}$ respectively. $\tau_{c}$ and $\mage$ are the biased ages while $\tau_{c}'$ and $\mcage$ are unbiased for the effects of secular acceleration. B1937+21 is likely to be significantly younger than its characteristic age. Associated errors that are more than 5\% of the most likely value are tabulated. Calculations are performed in double precision before rounding.}
\end{deluxetable}


\begin{deluxetable}{lccccc}
\label{tab:2}
\tablecolumns{4}
\tabletypesize{\scriptsize}
\singlespace
\tablewidth{0pt}
\tablecaption{Ages of millisecond pulsars with no proper motion measurements.
}
\tablehead{ 
\colhead{Pulsar} & 
\colhead{$\cage$ [Gyr]} &
\colhead{$\cage'$ [Gyr]} &
\colhead{ $\widetilde{\tau}$ [Gyr]} & 
\colhead{ $\widetilde{\tau}'$ [Gyr]} & 
\colhead{$\cage/ \widetilde{\tau}'$}
}
\startdata
        &    &      &  &         \\
 *  J0407+1607      &     5.15      &   5.64      &     5.06      &   5.55      &   0.93     \\
 *  J0609+2130      &     3.76      &   4.37      &     3.68      &   4.30      &   0.87     \\
 *  J0900\(-\)3144      &     3.59      &   5.11      &     3.47      &   4.99      &   0.72     \\
 *  J1038+0032      &     6.82      &   8.50      &     6.73      &   8.40      &   0.81     \\
 *  J1125\(-\)6014      &    10.39      &  16.37      &     8.89      &  14.00      &   0.74     \\
 *  J1157\(-\)5112      &     4.83      &   5.85      &     4.75      &   5.77      &   0.84     \\
 *  J1232\(-\)6501      &     1.72      &   1.75      &     1.67      &   1.69      &   1.02     \\
 *  J1420\(-\)5625      &     8.01      &  11.67      &     7.91      &  11.57      &   0.69     \\
 *  J1435\(-\)6100      &     6.05      &   6.92      &     5.92      &   6.79      &   0.89     \\
 *  J1439\(-\)5501      &     3.20      &   4.46      &     3.11      &   4.36      &   0.73     \\
 *  J1454\(-\)5846      &     0.88      &   0.89      &     0.81      &   0.83      &   1.06     \\
 *  J1528\(-\)3146      &     3.87      &   5.28      &     3.80      &   5.20      &   0.74     \\
 *  J1629\(-\)6902      &     9.51      &  18.16      &     9.24      &  17.66      &   0.54     \\
 *  J1721\(-\)2457      &     9.39      &  15.93      &     8.62      &  14.62      &   0.64     \\
 *  J1730\(-\)2304      &     6.37      &  42.92      &     6.24      &  42.27      &   0.15     \\
 *  J1732\(-\)5049      &     6.10      &   7.92      &     5.88      &   7.64      &   0.80     \\
 *  J1745\(-\)0952      &     3.23      &   3.56      &     3.14      &   3.46      &   0.93     \\
 *  J1751\(-\)2857      &     5.49      &   7.42      &     5.13      &   6.93      &   0.79     \\
 *  J1753\(-\)1914      &     0.49      &   0.50      &     0.44      &   0.45      &   1.10     \\
 *  J1753\(-\)2240      &     1.91      &   1.98      &     1.85      &   1.93      &   0.99     \\
 *  J1756\(-\)2251      &     0.44      &   0.45      &     0.38      &   0.38      &   1.16     \\
 *  J1757\(-\)5322      &     5.34      &   7.30      &     5.21      &   7.16      &   0.75     \\
 *  J1802\(-\)2124      &     2.78      &   2.95      &     2.68      &   2.84      &   0.98     \\
 *  J1804\(-\)2717      &     3.62      &   4.59      &     3.50      &   4.46      &   0.81     \\
 *  J1810\(-\)2005      &     3.44      &   3.66      &     3.36      &   3.57      &   0.96     \\
 *  J1841+0130      &     0.058      &   0.058      &     0.010$^{+0.013}_{-0.010}$      &   0.010$^{+0.013}_{-0.010}$      &   5.85$^{+\infty}_{-3.33}$     \\
 *  J1843\(-\)1113      &     3.05      &   3.41      &     2.15      &   2.41      &   1.27     \\
 *  J1853+1303      &     7.33      &  10.65      &     6.89      &  10.01      &   0.73     \\
 *  J1903+0327      &     1.81      &   1.85      &     1.42      &   1.45      &   1.25     \\
 *  J1904+0412      &    10.24      &  12.40      &    10.16      &  12.32      &   0.83     \\
 *  J1905+0400      &    12.34      &  33.12      &    11.47      &  30.81      &   0.40     \\
 *  J1910+1256      &     8.08      &  11.27      &     7.76      &  10.81      &   0.75     \\
 *  J1911+1347      &     4.29      &   5.24      &     4.09      &   4.99      &   0.86     \\
 *  J1918\(-\)0642      &     5.05      &   6.69      &     4.91      &   6.55      &   0.77     \\
 *  J2010\(-\)1323      &    17.17      & 185.03      &    16.54      & 178.24      &   0.10     \\
 *    J2033+17      &     8.57      &  14.86      &     8.33      &  14.44      &   0.59     \\
       \enddata
\tablecomments{We assume $v_{t}$=100\,km s$^{-1}$ for the Shklovskii correction in order to see the potential bias.  PSRs~J1022+1001, J1216\(-\)6410, J1829+2456 and J1933\(-\)6211 have $\dot{P}'<0$ for $v_{t}$=100\,km~s$^{-1}$. $\tau_{c}$ and $\mage$ are the biased ages while $\tau_{c}'$ and $\mcage$ are unbiased for the effects of secular acceleration. PSR J1841+0130 is likely to be significantly younger than its characteristic age. Associated errors that are more than 5\% of the most likely value are tabulated. Calculations are performed in double precision before rounding.} 
\end{deluxetable}

\end{document}